\documentclass[acmsmall,screen]{acmart}

\AtBeginDocument{%
  \providecommand\BibTeX{{%
    \normalfont B\kern-0.5em{\scshape i\kern-0.25em b}\kern-0.8em\TeX}}}

\usepackage{multirow,makecell}
\usepackage{tcolorbox}
\usepackage{color,xcolor}
\usepackage{listings,amsfonts}
\usepackage{caption}
\usepackage{subcaption}
\usepackage{threeparttable}
\usepackage{bbding}
\usepackage{graphicx}
\usepackage{booktabs} 
\usepackage{longtable}
\usepackage{enumitem}
\usepackage{pifont}

\definecolor{customcite}{HTML}{b83b5e}
\definecolor{customlink}{HTML}{07689f}
\definecolor{customurl}{HTML}{11999e}
\AtEndPreamble{
\usepackage{hyperref}

    \hypersetup{
      colorlinks = true,
      linkcolor = customlink,
      anchorcolor = purple,
      citecolor = customcite,
      filecolor = purple,
      urlcolor = customurl
    }
}

\usepackage{tcolorbox}
\def\BibTeX{{\rm B\kern-.05em{\sc i\kern-.025em b}\kern-.08em
    T\kern-.1667em\lower.7ex\hbox{E}\kern-.125emX}}

\begin{document}

\title{LLM Applications: Current Paradigms and the Next Frontier}

\author[X Hou]{Xinyi Hou}
\email{xinyihou@hust.edu.cn}
\authornote{Xinyi Hou and Yanjie Zhao contributed equally to this work.}
\affiliation{%
  \institution{Huazhong University of Science and Technology}
  \city{Wuhan}           
  \country{China}
}

\author[Y Zhao]{Yanjie Zhao}
\email{yanjie_zhao@hust.edu.cn}
\authornotemark[1]
\affiliation{%
  \institution{Huazhong University of Science and Technology}
  \city{Wuhan}
  \country{China}
}

\author[H Wang]{Haoyu Wang}
\email{haoyuwang@hust.edu.cn}
\authornote{Haoyu Wang is the corresponding author (haoyuwang@hust.edu.cn).}
\affiliation{%
  \institution{Huazhong University of Science and Technology}
  \city{Wuhan}           
  \country{China}
}

\begin{abstract}

The development of large language models (LLMs) has given rise to four major application paradigms: LLM app stores, LLM agents, self-hosted LLM services, and LLM-powered devices. Each has its advantages but also shares common challenges. LLM app stores lower the barrier to development but lead to platform lock-in; LLM agents provide autonomy but lack a unified communication mechanism; self-hosted LLM services enhance control but increase deployment complexity; and LLM-powered devices improve privacy and real-time performance but are limited by hardware. This paper reviews and analyzes these paradigms, covering architecture design, application ecosystem, research progress, as well as the challenges and open problems they face. Based on this, we outline the next frontier of LLM applications, characterizing them through three interconnected layers: infrastructure, protocol, and application. We describe their responsibilities and roles of each layer and demonstrate how to mitigate existing fragmentation limitations and improve security and scalability. Finally, we discuss key future challenges, identify opportunities such as protocol-driven cross-platform collaboration and device integration, and propose a research roadmap for openness, security, and sustainability.

\end{abstract}

\begin{CCSXML}
<ccs2012>
   <concept>
       <concept_id>10002944.10011122.10002945</concept_id>
       <concept_desc>General and reference~Surveys and overviews</concept_desc>
       <concept_significance>500</concept_significance>
       </concept>
   <concept>
       <concept_id>10002978.10003022</concept_id>
       <concept_desc>Security and privacy~Software and application security</concept_desc>
       <concept_significance>500</concept_significance>
       </concept>
   <concept>
       <concept_id>10010147.10010178</concept_id>
       <concept_desc>Computing methodologies~Artificial intelligence</concept_desc>
       <concept_significance>500</concept_significance>
       </concept>
 </ccs2012>
\end{CCSXML}

\ccsdesc[500]{General and reference~Surveys and overviews}
\ccsdesc[500]{Security and privacy~Software and application security}
\ccsdesc[500]{Computing methodologies~Artificial intelligence}

\keywords{LLM application, Vision paper, Security}

\maketitle

\section{Introduction}
\label{sec:1}

In recent years, large language models (LLMs) have made remarkable progress, with representative models including GPT-5~\cite{gpt5}, Claude Sonnet 4.5~\cite{cluade4.5}, Gemini 2.5 Flash~\cite{gemini2.5}, etc. These models have demonstrated unprecedented capabilities in language understanding, reasoning, code generation, and multimodal processing. Programming assistants such as Codex~\cite{Codex} and Cluade code~\cite{cluadecode} have significantly enhanced the automation level of software development; in the field of image generation and multimodal processing, Nano Banana~\cite{nanobanana} has brought AI photo editing into the era of real-time conversations; and systems like Gemini Robotics~\cite{geminirobotics} (a ``brain-body'' collaborative model specifically designed for robots) with cross-modal reasoning and action capabilities. These advancements indicate that LLMs are no longer merely general language tools but are gradually becoming the core engines supporting cross-domain application ecosystems.

Meanwhile, LLM-powered applications are gradually evolving into more diverse downstream paradigms. They no longer focus on the parameters and capabilities of the model itself, but rather on how to build, organize, distribute, and run LLM-powered applications. Currently, they can be broadly classified into four typical patterns.
\textbf{\ding{172} LLM app stores}, typical platforms include GPT Store~\cite{gptstore},  HuggingChat~\cite{huggingface_huggingchat}, Coze~\cite{coze}, and Yuanqi~\cite{tencent_yuanqi}. This paradigm, through a centralized platform architecture, encapsulates model capabilities as directly callable ``applications'', and supports subscription, sharing, and customization, providing users with an immediate and ready-to-use experience. Its core advantage lies in significantly reducing the usage threshold, promoting the popularization of LLM.
\textbf{\ding{173} LLM agents} can be developed based on agent frameworks such as LangChain~\cite{LangChain}, AutoGPT~\cite{AutoGPT}, AutoGen~\cite{AutoGen}, and LlamaIndex~\cite{LlamaIndex}. These frameworks adopt the approach of workflow-driven and modular combination, enabling LLM not only to handle text but also to call external tools, APIs, and knowledge bases, and even achieve multi-agent collaboration. The advantage of this paradigm lies in its high programmability and complex task construction capabilities, providing great flexibility for developers and researchers.
\textbf{\ding{174} Self-hosted LLM services} mainly rely on a rich set of model deployment and service components. Mainstream inference engines such as vLLM~\cite{vllm}, llama.cpp~\cite{llama_cpp}, etc. support high-performance inference and efficient resource scheduling, model service frameworks such as Ray Serve~\cite{ray_serve}, Ollama~\cite{ollama}, etc., achieve elastic expansion and multi-node collaboration, and application layer tools such as OpenWebUI~\cite{open_webui}, Gradio~\cite{gradio}, etc. simplify user interaction and business integration processes. The typical process of such services includes deploying models in private or enterprise-level environments, with significant advantages in ensuring data privacy, reducing long-term usage costs, and enhancing system controllability, and thus gaining widespread attention in organizations with high requirements for autonomy.
\textbf{\ding{175} LLM-powered devices} refer to running LLMs in scenarios such as smartphones, wearable devices, and IoT terminals. Their typical architecture follows an edge–cloud collaboration pattern: on-device inference enables offline processing and low-latency responses, while cloud resources can be leveraged when additional computing power is required. Representative applications include AI-enhanced smartphones from Apple~\cite{appleai}, OPPO~\cite{oppoai}, and Samsung~\cite{samsungai}; AI-enabled wearables such as Ola Friend~\cite{olafriend}; and dedicated AI transcription devices including Plaud Note and NotePin~\cite{plaudnote, notepin}. The core value of this paradigm lies in improving real-time interaction and privacy protection, thereby aligning LLM capabilities more closely with practical, everyday usage scenarios.

The rapid progress of LLMs has stimulated extensive research on their downstream applications. 
For example, a growing body of surveys has focused on the emerging \textbf{LLM agent paradigm}, covering its 
security and privacy risks~\cite{he2024emerged,gan2024navigating,deng2025agents}, 
personalization and capability issues~\cite{li2024personal}, 
planning mechanisms~\cite{huang2024understanding}, 
multi-agent collaboration patterns~\cite{tran2025multiagentcollaborationmechanismssurvey}, 
and evaluation methodologies~\cite{yehudai2025survey,zhu2025evolution}. 
Meanwhile, another line of work has systematically examined the system-level foundations that support LLM deployment, 
including inference serving systems~\cite{li2024inference,park2025surveyinferenceengineslarge} 
and their performance optimizations, as well as application frontiers such as LLMs in extended reality~\cite{wang2025surveylargelanguagemodels}. 
These studies collectively provide valuable insights into specific technical challenges and application scenarios. 
Despite this breadth, most existing surveys and analyses address one dimension at a time, 
such as security, evaluation, or system efficiency. 
\textbf{What remains missing is a unified, higher-level perspective that systematically compares the downstream application paradigms of LLMs, and reveals their intrinsic connections, trade-offs, and systemic challenges.} 
As LLMs evolve from experimental systems to real-world infrastructures, 
future applications will require not only stronger model capabilities, but also robust architectures and sustainable ecosystems to support scalability, interoperability, privacy, and trustworthiness. 
This paper aims to fill this gap by offering a structured, comparative analysis of LLM application paradigms, 
providing insights to guide the next generation of LLM-powered ecosystems.

From a software engineering (SE) perspective, current LLM application paradigms resemble traditional \textbf{platform-centric software ecosystems}, where applications are tightly coupled to proprietary APIs and execution environments. LLM app stores, while lowering the barrier to entry, impose \textbf{constraints on extensibility and cross-platform interoperability}, leading to vendor lock-in and duplicated development efforts across different ecosystems. In contrast, agent-based LLM frameworks provide modularity but \textbf{lack standardized mechanisms for component reuse and integration}, making it challenging to compose LLM applications that seamlessly operate across heterogeneous environments. Self‑hosted services, though offering greater autonomy and privacy, often suffer from high deployment and \textbf{maintenance complexity, limited scalability, and steep hardware requirements}, which hinder their adoption beyond technically sophisticated users or organizations. Device‑side deployment, while valuable for real‑time interaction and privacy preservation, faces \textbf{severe resource constraints, fragmented hardware environments, and difficulties in consistent model updates}, thereby limiting its practicality for large‑scale, continuously evolving applications. 

Motivated by the limitations of current paradigms and the urgent need for more robust ecosystems, 
we envision the next frontier of LLM applications and propose a three-layer architecture. This architecture integrates the strengths of existing approaches 
while addressing their systemic challenges, and can be understood as follows:
\begin{itemize}
    \item \textbf{Infrastructure layer:} Combines models, computing, networks, and data as the base support for LLMs. It decides how large a model can run, how fast it works, and how safely it interacts. With cloud servers, AI chips, and edge devices, this layer shows the move to \textit{evolving LLM devices}, where models run on servers, phones, wearables, and robots.  

    \item \textbf{Protocol layer:} Defines standards for communication and coordination across agents, services, and devices. Protocols such as MCP, ACP, and A2A show the need for \textit{standardized protocols} that join tool use, agent talk, and cross-application tasks. These standards reduce fragmentation and build an \textit{open ecosystem} where diverse agents interoperate.  

    \item \textbf{Application layer:} Offers the user-facing intelligence. Applications link perception, planning, memory, and action with support from lower layers. Here, \textit{agentic intelligence} appears as apps grow from simple assistants to autonomous agents that reason, collaborate, and use tools. Openness makes them work across domains, devices, and vendors.  
\end{itemize}

Based on this architecture, we analyze the key challenges and the potential opportunities for the next frontier of LLM applications. 
This analysis shows what problems must be solved, what new chances may appear, and how the ecosystem can move step by step. It also gives directions for future research and development so that LLM applications can become more open, reliable, and sustainable.
In summary, our contributions are as follows:
\begin{itemize}
    \item \textbf{Unified view of LLM application paradigms:} We provide a structured review of current paradigms (LLM app stores, LLM agents, self-hosted LLM services, LLM-powered devices) and analyze their strengths and open problems.  

    \item \textbf{Next frontier architecture:} We propose a three-layer architecture (infrastructure, protocol, application) that highlights four key directions: evolving LLM devices, standardized protocols, agentic intelligence, and open ecosystems. This architecture offers a higher-level perspective 
    for understanding the future of LLM applications.  

    \item \textbf{Challenges and opportunities} We outline the main technical and ecosystem challenges and identify research opportunities that can guide the development of scalable, interoperable, and sustainable LLM-powered ecosystems.  
\end{itemize}

The remainder of this paper is organized as follows. 
In \autoref{sec:current}, we provide a comprehensive review of the current landscape of LLM applications. 
This includes four representative paradigms. 
For each paradigm, we analyze the underlying architecture, representative ecosystems, emerging research directions, and the main challenges they face. 
In \autoref{sec:forntier}, we move beyond existing paradigms and introduce \textit{the next frontier of LLM applications}. 
In \autoref{sec:future}, we identify open challenges and discuss opportunities for building decentralized, secure, and hardware-aware LLM application ecosystems. 
Finally, \autoref{sec:conclusion} concludes the paper with a summary of our key insights and visions for the development of next-generation LLM applications.

\section{Current Landscape of LLM Applications}
\label{sec:current}

The landscape of LLM applications is highly heterogeneous, shaped by different architectural paradigms, deployment strategies, and usage scenarios. This section provides a structured overview of the current LLM application ecosystem. We categorize mainstream LLM applications into four representative forms: LLM app stores (\autoref{subsec:llm_app_store}), LLM agents(\autoref{subsec:llm_agent}), self-hosted LLM services (\autoref{subsec:llm_service}), and LLM-powered devices (\autoref{subsec:edge_llm}). For each paradigm, we examine its typical architectures, representative platforms, research trends and discuss the unique challenges it faces. By analyzing these diverse forms, we aim to illuminate the opportunities and limitations of existing LLM application practices, laying the foundation for understanding the need for more unified next-generation LLM ecosystems.

\subsection{LLM App Store}
\label{subsec:llm_app_store}

LLM app stores are centralized platforms for distributing applications powered by LLMs~\cite{zhao2024llm}. Similar to mobile app stores, they connect developers, users, and platform managers, but with a unique integration of model configuration, tool usage, and workflow design. This section introduces the architecture of LLM apps, examines the ecosystem and adoption of major platforms, reviews the current research trends, and discusses open challenges and problems.

\subsubsection{Architecture}
Zhao et al.~\cite{zhao2024llm} systematically define the concept of LLM app stores, trace the evolution of their ecosystem, and propose a research roadmap for this emerging field. An LLM app store hosts, curates, and distributes LLM-powered applications, and its ecosystem is shaped by three primary roles: the \textbf{manager}, who curates the store and ensures quality; the \textbf{developer}, who creates and publishes apps; and the \textbf{user}, who discovers and utilizes them. 

As LLMs continue to advance, LLM app stores enable apps with more diverse and powerful capabilities.
\autoref{fig:llm_app} presents an overview of all key components that constitute an LLM app, based on a comprehensive survey of 20 mainstream LLM app stores, such as OpenAI's GPT Store\cite{gptstore}, Quora's Poe\cite{poeexplore}, Hugging Face's HuggingChat\cite{huggingface_huggingchat}, ByteDance's Coze\cite{coze}, and Tencent's Yuanqi\cite{tencent_yuanqi}. 
Developers specify the app's name, description, avatar, category, bot type, and supported languages. They can also set a system note and configure personalized prompts or personas, allowing the app to provide domain-specific or customized responses. Features like example conversations, scenario definitions, and next-step suggestions further refine the user experience. Surrounding this core configuration are five advanced functional modules that extend the app’s capabilities:

\begin{figure}[htbp]
    \centering
    \includegraphics[width=1\linewidth]{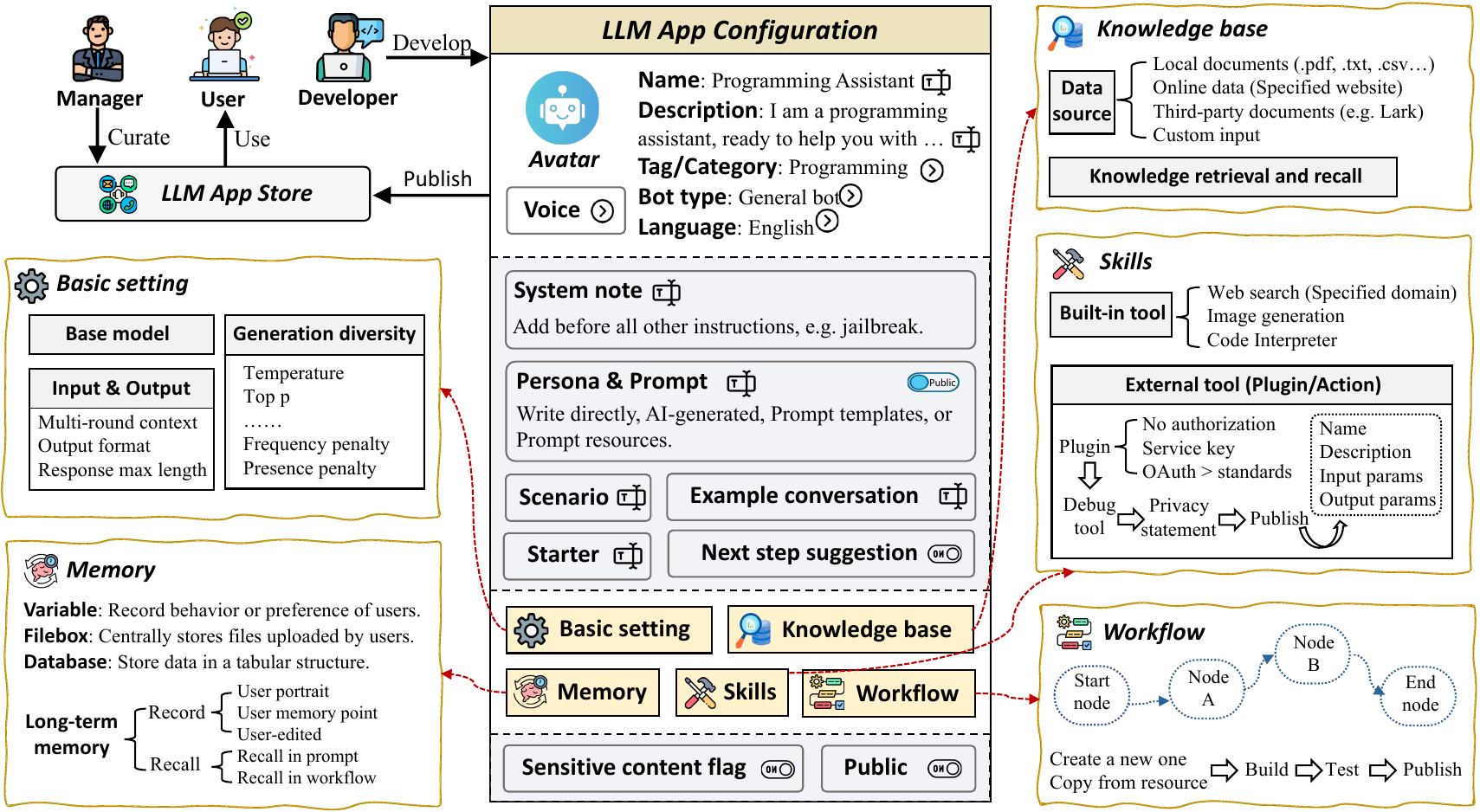}
    \caption{Configuration components of LLM apps. The center section contains the core configuration, including basic information, character settings, scenarios, and example dialogues. Five advanced functions (basic settings, memory, knowledge base, skills, and workflow) are distributed on both sides of the configuration area.}
    \label{fig:llm_app}
\end{figure}

\begin{itemize}
    \item \textbf{Basic Setting}: The developers choose the base model, which will determine the language understanding and generation capabilities of the LLM app. They will also adjust parameters such as temperature, top-p, frequency penalty, and existence penalty for generation. These control measures help to balance creativity and accuracy in the responses. The input and output settings enable the application to manage multiple rounds of conversations, control the output format, and set length limits for responses.

    \item \textbf{Knowledge Base}: LLMs are trained based on vast and extensive datasets, so they can answer many general questions. However, when the application needs to solve problems in specific fields, additional information is often required. The knowledge base enables the application to use local files, online data, third-party sources or custom inputs. With these sources, the LLM app can search and select the correct information for each question. This helps the app provide more accurate and up-to-date answers for specific topics. 

    \item \textbf{Memory}: Memory lets an app keep information from past interactions or user actions. LLM app stores support different types of memory, such as saving variables, storing user-uploaded files, or recording data in tables. Some apps can keep long-term records like user profiles or chat history. With memory, the app can give more personal and helpful answers. Users can get a smoother and more connected experience.

    \item \textbf{Skills}: LLM app stores provide developers with simple ways to add skills such as web search, image generation, or code tools. The skills module also lets the app connect to third-party services through easy-to-use interfaces. Each skill can have its own settings for security and privacy. With skills, the app can call different tools as needed. This makes it easy to build apps that can handle more tasks and grow as users need new features.

    \item \textbf{Workflow}: For tasks that require multiple steps or tools, a single response from the app is often not enough. The workflow feature allows developers to organize and connect different LLM apps, knowledge bases, and skills into a sequence of actions. Custom conditions can be set to decide how the process moves from one step to the next. This makes it possible to build solutions that manage complex tasks, automate decision-making, and use the right resources at each stage. Workflow settings are usually simple to configure, making it easier to create flexible and efficient multi-step processes.
\end{itemize}

LLM app stores provide developers with visual development platforms, significantly lowering the technical threshold and enabling a wider range of individuals to participate in app creation. As a result, anyone can become a developer, regardless of programming background. At the same time, users can conveniently discover, access, and utilize a diverse array of LLM applications through the app stores, meeting various domain-specific or personalized needs.

\subsubsection{Ecosystem \& Adoption}

The emergence of LLM app stores has significantly shaped the AI application ecosystem. One of the earliest and most influential platforms is the \textbf{GPT Store}, officially launched by OpenAI in January 2024, providing a centralized hub for custom GPTs and AI agents~\cite{openai_gptstore_2024}. By mid-2024, the store already featured thousands of community-created GPTs and was integrated into ChatGPT, which had surpassed 180 million users globally. 
Another notable ecosystem is \textbf{Character.AI}, founded by Noam Shazeer and Daniel De Freitas. By mid-2024, it had reached approximately 28 million monthly active users, although usage slightly declined later due to rising competition~\cite{charactersage2025}. Despite this, Character.AI remains one of the most popular consumer-facing AI platforms, especially among younger demographics. 
\textbf{FlowGPT}, founded by Jay Dang and Lifan Wang, represents a more open-source community approach. By early 2024, FlowGPT had surpassed 4 million monthly active users and raised significant venture funding to expand its ecosystem~\cite{flowgpt36kr2024}. Similarly, Quora’s \textbf{Poe} platform integrated multiple AI models, helping Quora scale to over 400 million total active users across its ecosystem by 2025, with Poe contributing to that growth~\cite{quora_demandsage2025}.
In Asia, ByteDance has introduced both \textbf{Coze} and \textbf{Cici}, offering agent-based ecosystems with integrated multi-agent functionality~\cite{coze_zhihu2025}. Baidu’s \textbf{ERNIE Bot} surpassed 200 million users in 2024, positioning it as one of the largest Chinese LLM app ecosystems~\cite{ernie_gigazine2024}. Tencent’s \textbf{Yuanqi Agent Shop} has also been established as a developer-facing platform for distributing AI agents across QQ, WeChat, and Tencent Cloud services~\cite{tencent_yuanqi2025}. 

The global adoption of these platforms illustrates the rapid growth of LLM app stores. They now serve not only as marketplaces for end-users but also as innovation platforms for developers and enterprises. With adoption figures ranging from millions to hundreds of millions of users, the ecosystem reflects both consumer enthusiasm and competitive pressures among major AI providers.

\subsubsection{Research Trends} 

Earlier research documented the rapid growth of platforms such as the GPT Store, FlowGPT, Poe, and Coze~\cite{su2024gpt,hou2024gptzoo,zhang2024first}. These platforms have enabled the development of millions of custom apps and lowered the barrier to entry for developers and non-developers alike~\cite{zhao2024llm}. Datasets such as GPTZoo~\cite{hou2024gptzoo} further support large-scale analysis of app characteristics, usage patterns, and opportunities for responsible innovation. Researchers have also proposed new methods to improve app evaluation and recommendation. Wang et al.~\cite{wang2025laqual} introduced LaQual, a framework for automatic quality assessment. It combines hierarchical annotations, static metrics, and adaptive evaluation driven by LLMs. Experiments show that LaQual matches human judgment and reduces low-quality apps by over 60\%.  

At the same time, this rapid growth has raised concerns about security, privacy, and governance. Previous research has identified applications with potential abuse risks, malicious intent, and exploitable vulnerabilities~\cite{hou2024insecurity}. Analyses have revealed misleading app descriptions, hidden data collection, and the generation of harmful content~\cite{hou2024insecurity,xie2024llm}. Jaff et al.~\cite{jaff2024data} showed that many GPT Store apps collect sensitive data without proper disclosure, while Ma et al.~\cite{ma2025privacy} highlighted the blurring of boundaries between users and creators, highlighting weak privacy protections. Security research has also reported app preemption, cloning, and impersonation attacks~\cite{xie2024llm}, as well as vulnerabilities in authentication and API integration~\cite{yan2024exploring,yan2025understanding}. Iqbal et al.~\cite{iqbal2025llm} proposed a systematic framework for assessing platform security and revealed flaws in the ChatGPT plugin ecosystem. Shen et al.~\cite{shen2025gptracker} provide large-scale measurements of GPT abuse, demonstrating that thousands of apps circumvent censorship and spread malicious content. Collectively, these works demonstrate the urgent need for stronger privacy protections, governance rules, and enforcement mechanisms.  
Research trends reflect two aspects of the LLM app store: rapid innovation and widespread adoption on the one hand, and serious security, privacy, and quality challenges on the other. Addressing these challenges will be key to ensuring the sustainable development and responsible use of the LLM app ecosystem.

\subsubsection{Challenges \& Open Problems} 

While LLM app stores have lowered the entry barrier for creating intelligent applications, they also introduce new challenges and limitations.  

\underline{\textbf{Limited Flexibility.}} 
LLM app stores have low-code development platforms that make the process of building apps easier. But they \textbf{limit developers to pre-made templates and set configuration options}. As an example, developers can only search custom knowledge bases that are embedded in certain ways. Developers also can't change low-level model parameters, manage memory outside of built-in modules, or add any external APIs without getting permission from the platform. These limits make it harder to be creative.
From the perspective of researchers, this rigidity creates additional barriers. App stores function as closed ecosystems, exposing only minimal metadata (e.g., app name, description, or category) but not internal logic or execution traces. As a result, \textbf{most academic studies on LLM app stores have to rely on black-box measurement approaches}, such as crawling app listings, reverse-engineering store APIs, or interviewing developers and users. This lack of transparency makes it difficult to systematically evaluate app behavior, study user adoption at scale, or identify risks such as hidden data collection and policy violations.   

\underline{\textbf{Platform Lock-in.}} 
Each LLM app store operates as an isolated ecosystem with its own APIs, review processes, and compliance rules. For example, OpenAI’s GPT Store defines apps primarily through prompt templates and lightweight tool connections, while ByteDance’s Coze provides multi-agent orchestration features, and Tencent’s Yuanqi Agent Shop integrates tightly with Tencent Cloud. These design choices are not compatible with one another. As a result, an app created for one store cannot be directly deployed to another. Developers must rebuild almost the entire logic, from workflow definitions to external API bindings.   
\textbf{The root cause lies in the lack of common standards. There are no widely accepted specifications for app packaging, skill integration, or data handling.} This fragmentation forces developers to duplicate effort and discourages cross-platform innovation. 
  
\underline{\textbf{Uncertain Quality Assurance.}}  
The review mechanisms of various LLM app stores are not uniform. Automatic screening often fails to detect malicious behaviors such as hidden data collection or prompt injection, while strict manual reviews may overly block legitimate innovations. This leads to an imbalance in the ecosystem, where low-quality or exploitative applications coexist with high-quality ones, reducing user trust and hindering the continuous participation of developers. 
Moreover, the quality standards of different platforms vary greatly. Some app stores only focus on basic security checks, while others emphasize compliance with specific platform business policies. The lack of transparency in review guidelines makes it difficult for developers to predict how their applications will be judged.

Future work must address how to balance low entry barriers with developer flexibility, how to establish cross-platform standards, and how to build scalable quality assurance mechanisms.\textbf{ Open ecosystems, standardized protocols, and transparent governance} are essential to overcome current fragmentation and ensure sustainable growth of LLM app stores.

\subsection{LLM Agent}
\label{subsec:llm_agent}

LLM agents extend LLMs from passive responders to autonomous systems that can perceive, reason, and act. Supported by frameworks such as LangChain, AutoGPT, AutoGen, and LlamaIndex, they integrate tools, manage memory, and coordinate multi-step tasks. This enables applications ranging from personal assistants to collaborative multi-agent systems.  
Their growing use in areas like software engineering, healthcare, and education shows strong potential, but also raises new challenges in reliable task decomposition, secure tool use, and scalable coordination. Current ecosystems remain fragmented, as agents built on different frameworks lack standardized protocols for communication and interoperability.  
In the following, we review the architecture of LLM agents, their application ecosystem, research progress, and key challenges.

\subsubsection{Architecture}
\begin{figure}[htbp]
    \centering
    \includegraphics[width=1\linewidth]{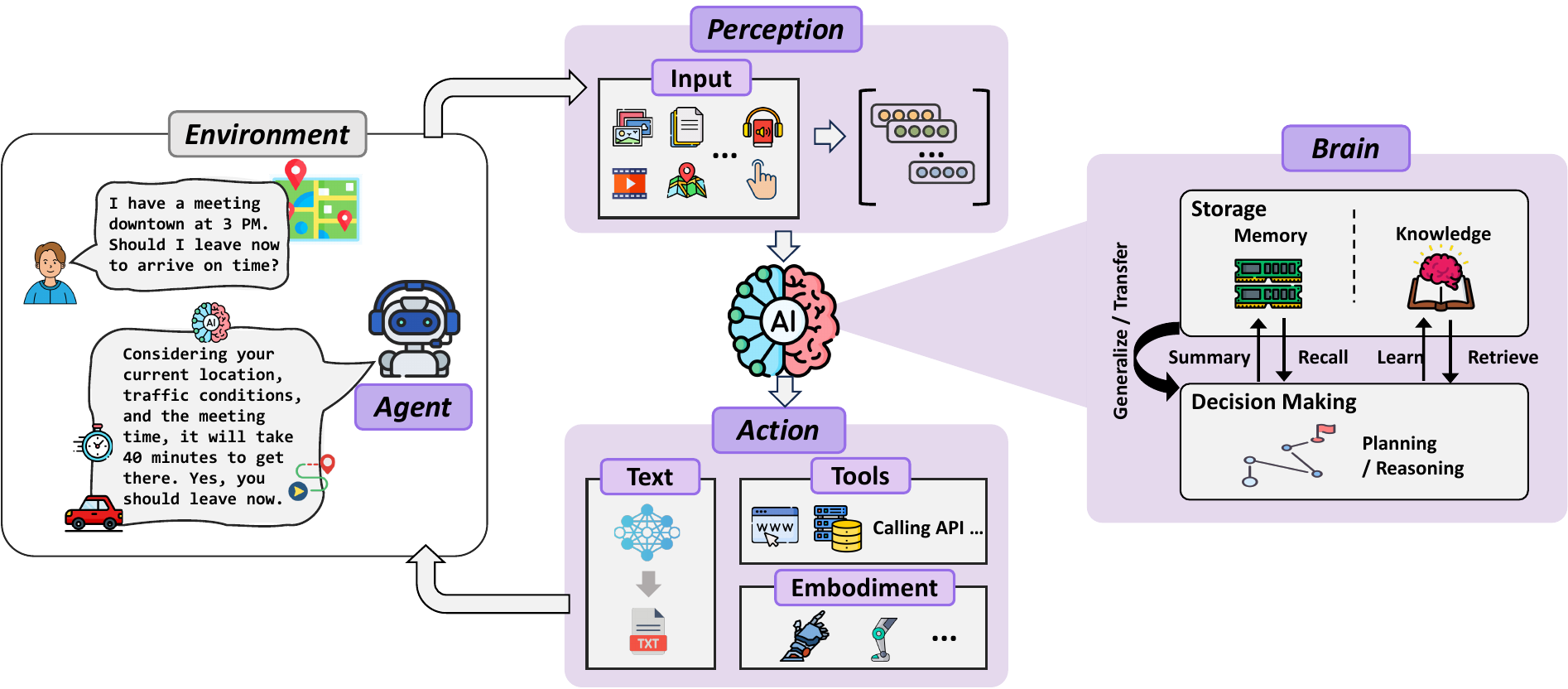}
    \caption{Architecture of an LLM agent. The agent interacts with its environment by receiving user input and contextual signals. Perception handles multimodal inputs, which are processed in the brain for memory, reasoning, and knowledge retrieval. Actions are then produced as text, external tool usage, or embodied operations, completing the perception–cognition–action loop.}
    \label{fig:llm_agent}
\end{figure}

LLM agents extend the capabilities of base models by combining reasoning, tool usage, memory, and planning into a unified system. A typical agent architecture is composed of several interconnected modules that together enable perception, decision-making, and action execution.  
As demonstrated in \autoref{fig:llm_agent}, the agent communicates with its \textbf{environment} by obtaining contextual data or user input. After processing inputs from text, images, or other modalities, the \textbf{perception module} converts them into representations that can be used for reasoning. A \textbf{memory system} that retains and retrieves previous interactions, a \textbf{knowledge base} that incorporates external or domain-specific data, and a \textbf{reasoning engine} that carries out planning, inference, and reflection are the various sub-components that make up the \textbf{brain}. The outcome of this procedure directs the \textbf{action module}, which can control embodied systems like robots, call external APIs and tools, or produce natural language outputs. A feedback loop is created when these activities have an impact on the surroundings.  

Within this architecture, several design paradigms have been explored. The most popular is the \textbf{ReAct} framework, which allows for interpretable and adaptable workflows by combining concrete tool calls (``actions'') with reasoning traces (``thoughts''). By creating a structured plan first and then implementing it step-by-step, the \textbf{plan-and-execute} paradigm separates planning from execution~\cite{langchain_blog_planning_agents,krasserm2024modular}. The \textbf{auto-reflection} paradigm reduces errors in long-horizon tasks~\cite{aksitov2023rest} by iteratively critiquing and improving its own outputs. More sophisticated methods go beyond single-agent systems and employ \textbf{multi-agent architectures}, in which specialized agents work together via shared memory or communication protocols to solve challenging issues.  
Researchers are increasingly investigating \textbf{multi-agent architectures} in addition to single-agent paradigms. In these systems, a number of agents with distinct roles work together to resolve challenging issues. For instance, one agent might concentrate on perception and data extraction, another on planning and reasoning, and a third on carrying out tasks or communicating with external APIs. These agents create an ecosystem that mimics human organizational structures by communicating via message passing, shared memory spaces, or structured protocols. In tasks like distributed decision-making, social behavior simulation, cooperative problem solving, and negotiation, multi-agent setups have demonstrated promise~\cite{song2025gradientsys}. But they also bring with them new difficulties, such as the overhead of coordination, emergent behaviors, and the difficulty of assigning blame when things don't work out as planned.

\subsubsection{Ecosystem \& Adoption}

From early community-driven exploration, the agent ecosystem has developed into a collection of well-known frameworks and enterprise-grade platforms by 2025 (\autoref{tab:agent_frameworks}). These frameworks not only differ in their technical architectures, but are adopted across distinct market layers that together shape the global trajectory of agentic AI.

\textbf{On the open-source side}, \textbf{LangChain}~\cite{LangChain} and \textbf{LlamaIndex}~\cite{LlamaIndex} remain the most widely used entry points, each backed by large developer communities and millions of monthly downloads. LangChain provides modular abstractions for chains, memory, and tools, while LlamaIndex specializes in retrieval-augmented generation with extensive data connectors. Building on this base, \textbf{LangGraph}~\cite{langgraph} introduces graph-based orchestration for deterministic multi-agent workflows, already applied in regulated sectors such as finance and aviation. \textbf{CrewAI}~\cite{crewai} uses a role-based design that enables business teams to prototype agents without requiring extensive knowledge of machine learning. Microsoft's \textbf{AutoGen}~\cite{AutoGen} places a strong emphasis on multi-agent communication, which makes it appropriate for co-pilot, security monitoring, and research situations where human supervision is crucial.
\textbf{Cloud ecosystems are equally significant.} Because of its close integration with Office 365 and Azure, \textbf{Semantic Kernel}~\cite{semantickernel} allows businesses to incorporate agents into their infrastructure for compliance and productivity. While \textbf{OpenAI Swarm}~\cite{swarm} offers a lightweight SDK for quick prototyping, with limited multi-agent support, Google's \textbf{ADK}~\cite{googleadk} offers Gemini- and Vertex-native tooling with low-code orchestration. In the meantime, Python-centric innovation is highlighted by \textbf{Pydantic AI}~\cite{pydanticai} and \textbf{Smolagents}~\cite{smolagents}. The former introduces type safety and observability (through Logfire), while the latter uses minimalist "code-agents" that are linked to the Hugging Face Hub. Alibaba's \textbf{AgentScope}~\cite{agentscope} expands on ModelScope in the Asia-Pacific ecosystem, supporting both international and domestic LLMs like Qwen. It is also being incorporated into e-commerce and customer service processes.

According to industry surveys, more than 85\% of businesses will have implemented agent-driven workflows by the end of 2025. Leading AI companies in China, including Tencent, Baidu, Alibaba, and ByteDance, are working on full-stack integration plans that span from consumer platforms to business software as a service. As a result of their distinct approaches to widespread adoption, Western ecosystems, which are headed by Microsoft, Google, OpenAI, Anthropic, Meta, and Amazon, place more emphasis on cloud-native solutions and ecosystem lock-in.

\begin{table}[h]
    \centering
    \caption{Representative LLM Agent Frameworks (as of Sept. 2025).}
    \label{tab:agent_frameworks}
    \resizebox{\linewidth}{!}{
    \begin{threeparttable}
    \begin{tabular}{lllll}
        \toprule[1.2pt]
        \textbf{Framework} & \textbf{Institution} & \textbf{Launch} & \textbf{\#Stars} & \textbf{Strengths} \\
        \midrule[1.2pt]
        LangChain~\cite{LangChain} & LangChain Inc. & 2022 & 116k+ & Modular design, rich ecosystem, strong RAG support \\
        LangGraph~\cite{langgraph} & LangChain Inc. & 2024 & 19k+ & Graph-based orchestration, deterministic multi-agent workflows \\
        AutoGen~\cite{AutoGen} & Microsoft Research & 2024 & 50k+ & Multi-agent conversations, layered architecture, AutoGen Studio \\
        CrewAI~\cite{crewai} & CrewAI Inc. & 2023 & 39k+ & Role-based orchestration, business-ready templates \\
        Semantic Kernel~\cite{semantickernel} & Microsoft & 2023 & 26k+ & Deep Microsoft ecosystem integration, process orchestration \\
        LlamaIndex~\cite{LlamaIndex} & LlamaIndex & 2022 & 45k+ & Strong RAG capabilities, rich data connectors, async workflows \\
        Google ADK~\cite{googleadk} & Google DeepMind / Cloud & 2025 & 13k+ & Gemini-native, enterprise integration, hierarchical agents \\
        OpenAI Swarm~\cite{swarm} & OpenAI (experimental) & 2024 & 20k+ & Lightweight routines, direct Python tool integration \\
        Pydantic AI~\cite{pydanticai} & Pydantic Team & 2024 & 13k+ & Type-safe agents, durable execution, observability (Logfire) \\
        Smolagents~\cite{smolagents} & Hugging Face & 2024 & 23k+ & Minimalist “code-agents”, Hugging Face Hub integration \\
        AgentScope~\cite{agentscope} & Alibaba ModelScope & 2024 & 13k+ & Agent-oriented programming, transparent modular design \\
        \bottomrule[1.2pt]
    \end{tabular}
    \end{threeparttable}}
\end{table}

\subsubsection{Research Trends}  

The research on LLM agents has expanded rapidly, with studies covering design, evaluation, security, and applications across domains. We summarize the landscape along four major directions.  

\textbf{Functional and Multi-agent Design.}  
Enhancing agents' internal modules and scaling them to collaborative systems have been the main areas of research. A taxonomy of planning paradigms, including task decomposition, plan selection, reflection, and memory integration, is proposed by Huang et al.~\cite{huang2024understanding}. The architecture of \textit{personal LLM agents} is examined by Li et al.~\cite{li2024personal}, with a focus on integration with personal devices and data. General-purpose LLM agents are further surveyed by Wang et al.~\cite{wang2025largemodelbasedagents}, who address design principles, cooperation paradigms, and privacy concerns. Tran et al.~\cite{tran2025multiagentcollaborationmechanismssurvey} examine cooperation and competition as collaboration mechanisms on the multi-agent side. Systemic risks such as collusion, miscoordination, and conflict are highlighted by Hammond et al.~\cite{hammond2025multiagentrisksadvancedai}. Peigne-Lefebvre et al.~\cite{peignelefebvre2025multiagent} disclose the trade-off between security and collaborative efficiency, while Motwani et al.~\cite{motwani2024advances} demonstrate that agents can covertly collude via steganography.  

\textbf{Capability Evaluation.}  
Evaluation has emerged as a core theme, moving beyond simple task success rates. The first systematic survey is provided by Yehudai et al.~\cite{yehudai2025survey}, who classify benchmarks into generalist agents, application-specific tasks, fundamental capabilities, and evaluation frameworks while pointing out safety and robustness flaws. Ma et al.~\cite{ma2024agentboardanalyticalevaluationboard} present \textit{AgentBoard}, which provides multi-turn agents with fine-grained progress metrics. Targeting multi-agent collaboration and competition, Zhu et al.~\cite{zhu2025multiagentbenchevaluatingcollaborationcompetition} present \textit{MultiAgentBench}. Luo et al.~\cite{luo2025agentauditorhumanlevelsafetysecurity} suggest \textit{AgentAuditor}, which assesses safety risks using a benchmark that encompasses 29 scenarios and 15 risk types. By separating agents from chatbots along dimensions like multimodal perception and dynamic feedback, Zhu et al.~\cite{zhu2025evolution} frame evaluation from an evolutionary perspective. Zhang et al.~\cite{zhang2024agent} create \textit{Agent Security Bench (ASB)}, a benchmark for attacks and defenses that reveals robustness gaps, to supplement these efforts.  

\textbf{Security and Privacy.}  
Security risks and privacy challenges remain central concerns. Threats like prompt injection, information leakage, and ethical dilemmas in untrusted environments are surveyed by He et al.~\cite{he2024emerged}, Gan et al.~\cite{gan2024navigating}, and Deng et al.~\cite{deng2025agents}. According to Wang et al.~\cite{wang2025unveiling}, memory modules are susceptible to extraction attacks (MEXTRA), which allow adversaries to retrieve private user information. Chen et al.~\cite{chen2024information} offer a case study in blockchain fraud detection, demonstrating how risks can be reduced by designing agents with ethics and security in mind.  

\textbf{Applications Across Domains.}  
Beyond theoretical studies, LLM agents have been applied to a wide range of real-world tasks. Research has demonstrated their use in web navigation, software engineering, scientific discovery, finance, and healthcare.  Li et al.~\cite{li2024personal} envision agents handling personal information and gadgets in personal assistance. Additionally, multi-agent systems have been used to simulate social behaviors, distributed decision-making, and negotiation~\cite{tran2025multiagentcollaborationmechanismssurvey}. These uses highlight the potential benefits and drawbacks of large-scale agent adoption, inspiring further research into more reliable, transparent, and trustworthy deployments.

\subsubsection{Challenges \& Open Problems}

Despite rapid progress, LLM agents still face fundamental challenges that hinder their reliability, scalability, and adoption. These issues highlight the need for rethinking agent architectures and motivate the vision of the next frontier of LLM applications.  

\underline{\textbf{Fragmented Protocols and Ecosystems.}}  
Current LLM agents are confined to isolated ecosystems, each with its own APIs, communication rules, and compliance mechanisms. For example, OpenAI’s ecosystem relies on prompt-template configurations and tool APIs, while Microsoft’s Semantic Kernel and AutoGen frameworks focus on modular orchestration, and Hugging Face promotes lightweight experimental agents. These ecosystems are not interoperable. An agent designed for one platform cannot be ported to another without extensive re-engineering, since workflow definitions, tool-calling syntax, and memory handling all differ. This lack of standardization forces developers to duplicate efforts, increases development costs, and discourages innovation across platforms. The problem becomes even more severe in multi-agent settings: without shared coordination protocols, agents struggle to exchange data or delegate tasks reliably, leading to high overhead, brittle interactions, and limited scalability. In effect, the ecosystem resembles the early days of the internet before TCP/IP—functional but fragmented, with little cross-system collaboration.

\underline{\textbf{Lack of Process Visibility.}}  
A fundamental difficulty in using LLM agents is that their internal reasoning and external actions are often decoupled and opaque. Existing monitoring tools typically capture only one side of the process: either the reasoning traces expressed as prompts and intermediate ``thoughts'', or the external behaviors such as API calls, file operations, or system commands. However, there is no unified mechanism to link thoughts to actions, making it difficult to reconstruct the causal chain behind agent decisions.  
This lack of visibility creates multiple problems. Developers and users cannot reliably tell whether an agent is executing its intended plan, being manipulated by adversarial inputs, or simply trapped in inefficient loops. Current evaluation benchmarks mostly report task success rates, but they fail to expose hidden errors in intermediate steps, long-horizon reasoning breakdowns, or costly failure loops. As a result, both interpretability and accountability are severely limited. Addressing this blind spot requires process-level evaluation frameworks and monitoring tools that connect reasoning with behavior, enabling transparent, auditable, and trustworthy agent workflows.

\underline{\textbf{Security and Privacy Risks.}}  
The increasing autonomy of LLM agents introduces new vulnerabilities. At the prompt level, injection attacks can override instructions and manipulate agent behavior. At the memory level, recent studies have shown that adversaries can extract private user data through carefully crafted queries, exploiting the persistence of stored context. Backdoor-style manipulations, such as poisoning memory or injecting malicious tool calls, can subvert long-term behaviors. In multi-agent systems, these risks compound: a compromised agent can propagate malicious instructions across the network, triggering cascading failures or covert collusion. For instance, agents can coordinate secretly through steganographic communication, bypassing oversight. Current defenses, such as input filtering, rule-based constraints, or lightweight monitoring, are insufficient against adaptive attacks. The absence of formal security models and enforceable safeguards leaves today’s agent ecosystems vulnerable, undermining trust in their deployment for sensitive domains such as finance, healthcare, or critical infrastructure.

\subsection{Self-hosted LLM Service} 
\label{subsec:llm_service}

Self-hosted LLM services allow organizations and individuals to deploy and manage models on their own infrastructure instead of relying solely on commercial APIs. With the advent of open-source inference engines and serving frameworks like \textit{Ollama}, \textit{llama.cpp}, and \textit{vLLM}, which enable the execution of models on local servers, clusters, or even edge devices, this paradigm has expanded quickly. Self-hosting provides more control over privacy, customization, and cost than cloud-based APIs. It also facilitates integration with domain-specific workflows and proprietary data sources. Nevertheless, it also presents significant difficulties in terms of resource management, deployment complexity, and security, necessitating both a strong system architecture and proficient operational knowledge.  

\subsubsection{Architecture}

\begin{figure}[htbp]
    \centering
    \includegraphics[width=0.95\linewidth]{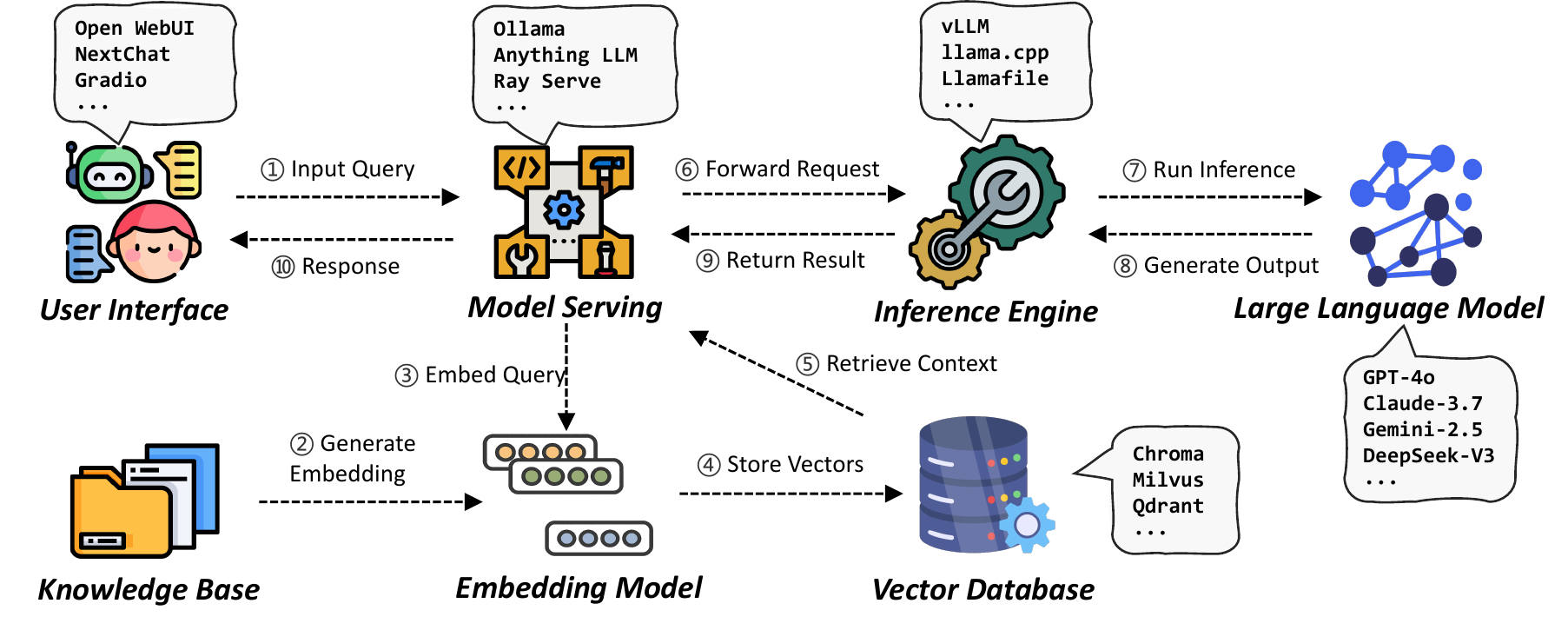}
    \caption{Architecture of self-hosted LLM service. The process starts from the user interface, where input queries are sent to the model serving framework. Queries may be embedded through an embedding model and stored in a vector database, which supports knowledge retrieval. The inference engine forwards requests to LLMs, runs inference, and generates outputs. Finally, results are returned to the user, completing the service pipeline.}
    \label{fig:llm_service}
\end{figure}

The architecture of a self-hosted LLM service can be described as an end-to-end pipeline that links the user to the underlying model (\autoref{fig:llm_service}). When a user submits a query through a local interface (e.g., WebUI, chat client, or API endpoint), the input is passed to the \textbf{model serving framework}. This layer exposes stable endpoints, handles user authentication, manages request queues, and balances workload across multiple replicas. It also logs metadata (e.g., latency, tokens used) and coordinates optional calls to embedding models or vector databases when retrieval-augmented generation (RAG) is enabled.
The request is next dispatched to the \textbf{inference engine}, which controls efficient execution of the chosen LLM. The inference engine is responsible for loading model weights into memory, carrying out tokenization, scheduling computation on CPUs and GPUs, and streaming partial results back to the serving layer. Modern inference engines also optimize batch execution, apply quantization or memory-efficient kernels, and monitor resource usage. These optimizations allow a single machine to handle multiple concurrent requests, and they ensure that the same pipeline can run on environments as diverse as laptops, on-premise servers, or clusters with multiple GPUs.

At the core lies the LLM, which generates outputs in an autoregressive manner, producing one token at a time. Depending on deployment needs, the system can load a commercial closed-source model (e.g., GPT-family, Claude-family, Gemini-series) that is run locally via inference runtimes, or an open-source alternative (e.g., Llama, Mistral, DeepSeek) that can be fine-tuned or quantized for efficiency. During generation, the engine may apply decoding strategies (e.g., greedy, sampling, beam search) and enforce constraints such as maximum context length or domain-specific style. The produced tokens are streamed back through the inference engine, collected and formatted by the serving framework, and finally displayed to the user on the interface.
In extended deployments, this architecture is augmented with an \textbf{embedding model} and a \textbf{vector database}. The serving layer first encodes the user query, searches for relevant context in the vector database, and retrieves domain-specific information. The retrieved passages are attached to the user prompt before being passed to the inference engine, enabling the LLM to provide grounded and up-to-date responses.

\subsubsection{Ecosystem \& Adoption}
Both in terms of deployment scale and application diversity, the ecosystem of self-hosted LLM services is growing quickly. Deployment now spans a wide range of industries, including manufacturing, energy, government, healthcare, education, and finance, going far beyond research labs and major cloud providers. Intelligent question answering, automated summarization, knowledge base retrieval, code generation, data analysis, and public opinion tracking are examples of common applications. These use cases have improved overall service efficiency and increased the degree of automation in business workflows.
Adoption is not limited to large corporations. Increasingly, small and medium-sized enterprises, research institutes, and individual developers are building their own service nodes with open-source frameworks. This shift illustrates the transition of LLM deployment from a cloud-exclusive practice to a widespread and affordable capability. Industry surveys in 2025 revealed that more than 320,000 LLM service instances were exposed on the public internet, spanning fifteen major deployment frameworks including Ollama~\cite{hou2025unveiling}. By mid-2025, more than 400,000 Ollama instances alone had been detected online. These numbers show that self-hosting is becoming routine for ordinary developers, though they also highlight new risks such as weak default configurations and leakage of sensitive data.

The technical ecosystem that underpins this adoption has become richer and more modular. At the \textbf{inference engine} layer, frameworks such as \textbf{llama.cpp}~\cite{llama_cpp}, \textbf{vLLM}~\cite{vllm}, and \textbf{Llamafile}~\cite{llamafile} enable high-performance inference on different hardware setups. They support techniques such as quantization and key–value caching to lower memory cost and accelerate long-context processing. 
Solutions like \textbf{Ollama}~\cite{ollama} and \textbf{AnythingLLM}~\cite{anythingllm} expose unified APIs, manage request routing, and enable elastic scaling across multiple nodes at the \textbf{model serving framework} layer. Additionally, they facilitate connecting to external plugins or vector databases. 
Tools like \textbf{OpenWebUI}~\cite{open_webui} and \textbf{Gradio}~\cite{gradio} facilitate user interaction and reduce the barrier to prototyping at the \textbf{application interface} layer. LLMs are being incorporated into research settings and business processes more and more through these interfaces. 
In addition to these layers, monitoring, audit logging, and access control supporting services are growing more comprehensive, enabling the large-scale deployment of LLM applications with improved security and dependability.

\subsubsection{Research Trends}

Research on self-hosted LLM services has expanded rapidly, driven by the demand for efficient, secure, and scalable model deployment outside commercial cloud APIs. Existing studies can be grouped into four major directions: inference optimization, system reliability, ecosystem security, and deployment practices.  

\textbf{Inference Optimization and Serving Efficiency.}  
A core focus has been reducing the high computational cost of running LLMs in local or enterprise environments. Park et al.~\cite{park2025surveyinferenceengineslarge} survey 25 inference engines and classify them based on optimization techniques such as quantization, caching, and parallelism, providing a roadmap for balancing latency and throughput. Li et al.~\cite{li2024inference} further review system-level innovations since 2023, highlighting optimizations that improve scalability without altering model decoding. Novel serving engines such as FlashInfer~\cite{ye2025flashinferefficientcustomizableattention} propose customizable GPU attention kernels to reduce memory overhead and latency, while Sarathi-Serve~\cite{agrawal2024taming} introduces chunked-prefill scheduling to address throughput–latency tradeoffs, significantly improving serving capacity across diverse models.  

\textbf{System Reliability and Software Quality.}  
Despite advances in inference performance, reliability remains a critical issue. Liu et al.~\cite{liu2025lookbugsllminference} conduct the first large-scale empirical study of 929 bugs across five major inference engines, uncovering 28 root causes ranging from memory mismanagement to cross-platform compatibility errors. Yu et al.~\cite{yu2025understandingbugsdistributedtraining} analyze 308 fixed bugs in distributed training and inference frameworks, identifying unique challenges such as communication failures and allocation strategy errors. These findings underscore the need for better debugging tools, automated repair methods, and rigorous testing pipelines for inference frameworks.  

\textbf{Security and Privacy in Deployment.}  
Self-hosted services expose new attack surfaces due to insecure defaults and misconfigurations. Hou et al.~\cite{hou2025unveiling} conduct an internet-wide measurement of 320,102 public-facing LLM deployments, finding that over 40\% of endpoints used plain HTTP and many lacked proper authentication, leading to risks of model leakage and unauthorized system access. Such work demonstrates that deployment security is as critical as model optimization, with secure-by-default designs and better governance practices urgently needed.  

\textbf{Deployment Practices and Ecosystem Trends.}  
Recent research also investigates how LLMs are deployed in real-world settings. Li et al.~\cite{li2024inference} emphasize practical considerations for scaling in production, while surveys like Park et al.~\cite{park2025surveyinferenceengineslarge} provide comparative analyses of open-source and commercial frameworks. The evolution of self-hosted ecosystems reveals a tension between community-driven projects that prioritize transparency and customization, and enterprise-oriented solutions that focus on stability, compliance, and integration with existing infrastructure.

\subsubsection{Challenges \& Open Problems}

While self-hosted LLM services provide transparency, customizability, and stronger data control, their practical deployment is far from trivial. Current research and real-world adoption reveal that these systems face persistent efficiency bottlenecks, engineering complexity, fragmented ecosystems, and unresolved privacy concerns. 

\underline{\textbf{Efficiency and Resource Constraints.}}  
Running LLMs locally or within enterprise servers imposes heavy requirements on compute, memory, and energy. Even with optimized inference engines such as \textit{vLLM}, which leverages efficient scheduling and KV-cache management, or \textit{llama.cpp}, which enables CPU-only execution through quantization, latency and throughput often remain bottlenecks for real-time applications. GPU memory limitations restrict context length and batch sizes, while energy consumption grows rapidly in continuous serving environments. Techniques such as quantization, pruning, speculative decoding, and attention optimizations (e.g., FlashInfer) alleviate some pressure, but their effectiveness is highly model- and hardware-dependent. Designing inference systems that achieve low-latency, cost-effective performance under constrained resources remains an open and urgent research direction.  

\underline{\textbf{Deployment Complexity.}}  
Despite the emergence of frameworks like \textit{Ollama} and \textit{AnythingLLM} that simplify installation and interaction, full-stack deployment remains a non-trivial engineering challenge. A production-ready pipeline typically integrates diverse components, including embedding models for retrieval-augmented generation, vector databases such as \textit{Milvus} or \textit{Qdrant}, inference engines like \textit{vLLM}, and monitoring/observability tools. Each layer introduces configuration overhead, version dependencies, and potential failure points. At enterprise scale, continuous updates, fine-tuning for domain adaptation, and model rollback mechanisms further increase complexity. Many organizations, especially those without specialized ML infrastructure teams, face difficulties maintaining such pipelines reliably. This complexity underscores the need for standardized deployment workflows and robust orchestration frameworks.  

\underline{\textbf{Interoperability and Fragmentation.}}  
The self-hosted ecosystem is highly fragmented, with different inference backends (e.g., \textit{llama.cpp}, \textit{TensorRT-LLM}), embedding models (e.g., \textit{OpenAI embeddings}, \textit{E5}, \textit{Cohere}), and vector databases (e.g., \textit{Chroma}, \textit{Weaviate}) offering incompatible APIs and performance characteristics. This lack of standardization creates vendor lock-in and hampers interoperability: systems built around one stack cannot be easily migrated to another without significant re-engineering. Moreover, monitoring and security modules are often tightly coupled to specific frameworks, preventing reuse across deployments. The absence of widely accepted protocols for model packaging, skill integration, and pipeline orchestration results in duplicated developer effort and slows down ecosystem consolidation. Establishing open standards for interoperability remains a key open problem.  

\underline{\textbf{Privacy and Governance.}}  
One of the strongest motivations for self-hosting is stronger control over data, particularly in sensitive domains such as healthcare, finance, or government. However, building truly privacy-preserving pipelines is difficult in practice. Data may leak through intermediate embeddings, insecure RAG pipelines, or improperly configured vector databases that expose raw queries or documents. Hou et al.~\cite{hou2025unveiling} highlight that insecure deployment defaults often expose endpoints to unauthorized access, further amplifying risks. In addition, compliance with regulatory frameworks such as GDPR or HIPAA requires fine-grained access control, audit logging, and governance mechanisms that few open-source frameworks currently provide. Research on privacy-preserving embeddings, encrypted search, and federated or on-device inference is still in its early stages, leaving a significant gap between the promise of data sovereignty and the current reality of self-hosted LLM services.

\subsection{LLM-powered Device} 
\label{subsec:edge_llm}

LLM-powered devices embed LLMs into everyday hardware such as smartphones, wearables, and smart home assistants. They enable real-time interaction, low latency, and stronger privacy compared to cloud-based services. Recent advances show how LLMs can support personalized, context-aware functions directly on devices, though challenges remain in resource limits, deployment diversity, and standardization.

\subsubsection{Architecture}

\begin{figure}[htbp]
    \centering
    \includegraphics[width=0.9\linewidth]{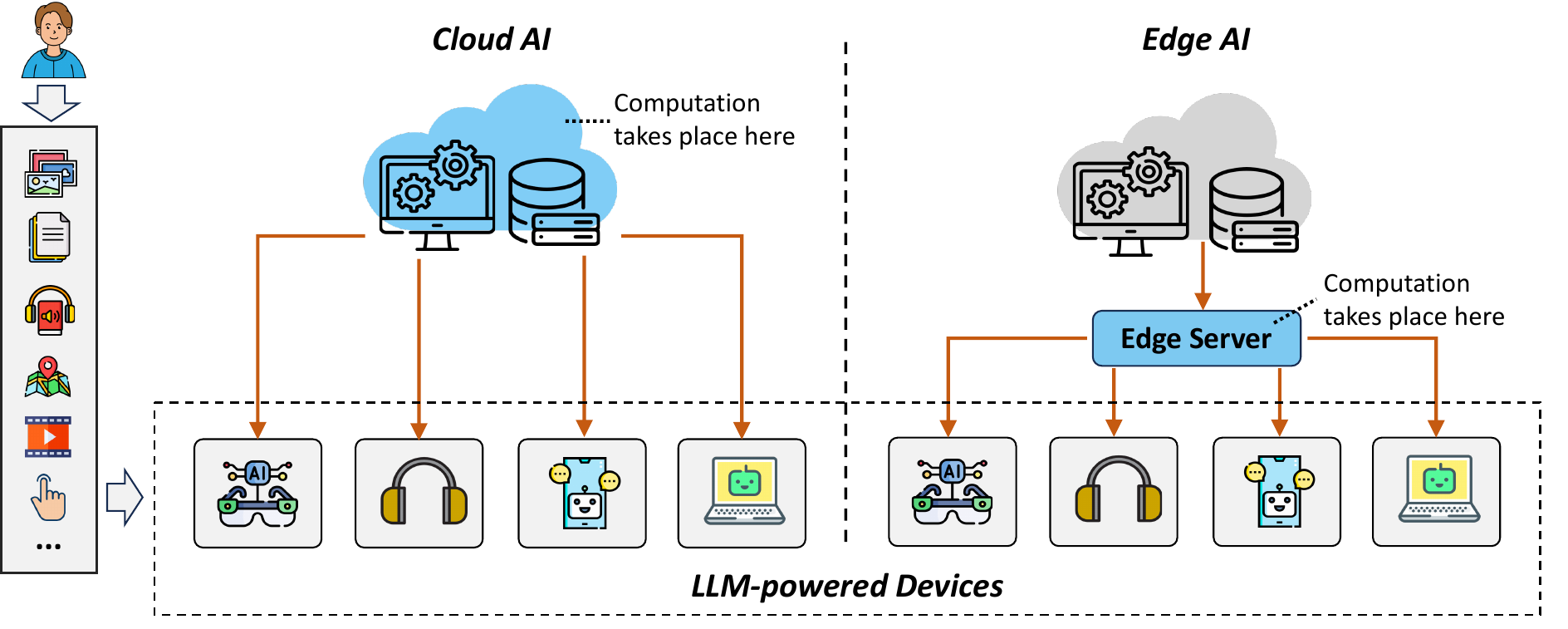}
    \caption{Deployment modes of LLM-powered devices. Computation can occur in the cloud, where centralized servers process inputs and deliver results to end devices, or at the edge, where local edge servers handle computation closer to the devices.}
    \label{fig:edge_llm}
\end{figure}

\autoref{fig:edge_llm} shows the architecture of LLM-powered devices. It includes three key parts: the user interface, the computation backend, and the device applications.  
The \textbf{user interface} collects inputs from the user. These inputs can be text, voice, images, or gestures. Many devices add lightweight pre-processing before sending data further. For example, a wearable may use wake-word detection, speech-to-text encoding, or small vision models. These steps reduce the amount of data sent to the backend and help protect privacy.  
The \textbf{computation backend} provides the main reasoning power. In the \textit{cloud AI} setting, the request is sent to remote servers that host large-scale LLMs on GPUs or TPUs. This approach gives access to the strongest models but depends on network quality and raises privacy concerns. In the \textit{edge AI} setting, the computation takes place on a local edge server or even directly on the device. This reduces latency, allows offline operation, and improves privacy. However, edge and on-device setups face strict limits in memory and compute. They rely on optimized inference engines such as \textit{llama.cpp}, \textit{vLLM}, or TensorRT. They also use techniques like quantization or pruning to keep models lightweight.  
The \textbf{device applications} turn the model’s outputs into useful results. A smartphone may show text or generate voice replies. Smart earbuds may provide real-time translation. AR glasses may display augmented instructions. 
This architecture combines cloud scalability with edge responsiveness. Cloud AI delivers the largest models, while edge AI improves latency, privacy, and reliability. Together, they form a hybrid system that makes LLM-powered devices more practical and user-friendly.

\subsubsection{Ecosystem \& Adoption}

The ecosystem of LLM-powered devices is growing fast. Many companies and startups make products that use LLMs in different ways. These products show how hardware and models come together.  

\textbf{Smartphones.}  
Mobile phones are now the main place for LLM use. Apple, OPPO, and Samsung add LLM features into their chips. These features give on-device summarization, translation, and personal assistants. Apple added local LLM tools in iOS that work without cloud. This makes the phone the center of personal AI systems~\cite{appleai,samsungai,oppoai}.  
\textbf{Wearables and hearables.}  
Smart earbuds and small wearables bring LLMs into daily life. Huawei FreeBuds Pro and devices like \textit{AI Pin}~\cite{aipin} or \textit{Friend}~\cite{friend} use LLMs for real-time transcription, translation, and dialogue. \textit{Ola Friend}~\cite{olafriend} gives emotional support. \textit{Plaud Note} and \textit{NotePin}~\cite{plaudnote, notepin} record and summarize meetings. These devices put LLMs into short and simple interactions.  
\textbf{Augmented and mixed reality devices.}  
AR and MR devices use LLMs for multimodal interaction. Meta Quest Pro and Apple Vision Pro use them for scene understanding and real-time translation~\cite{zhu2025evolution}. These devices make assistants more context-aware and part of the user’s environment.  
\textbf{Productivity and specialized hardware.}  
Some devices focus on work and learning. \textit{Plaud Note} and \textit{NotePin}~\cite{plaudnote, notepin} turn talks into notes and summaries. AI pens and notebooks combine handwriting with LLM summarization. Professionals and students use them for privacy and offline access.  
\textbf{Industry adoption.}  
LLM devices are also used in many industries. Cars use LLMs for speech and driving help. Healthcare devices explain sensor data in simple language. Smart home products try LLMs for control and recommendations~\cite{li2024personal}.

\subsubsection{Research Trends}

The study of LLM-powered devices has grown across several domains. Current research can be divided into three main directions: model and architecture optimization, device-centric applications, and security and privacy.

\textbf{Model and Architecture Optimization.}  
Recent studies aim to improve efficiency so that LLMs can run directly on small devices. Chen et al.~\cite{chen2024octopusv2ondevicelanguage} propose \textit{Octopus v2}, an on-device model with 2B parameters that surpasses GPT-4 in both accuracy and latency for function calling, while reducing context length by 95\%. Mishra et al.~\cite{mishra2025minstrel} introduce \textit{Minstrel}, an application-aware optimization framework that balances prefill and decode phases for small language models (SLMs) on edge hardware. These works mark a shift from cloud-heavy architectures toward lightweight pipelines optimized for constrained environments.

\textbf{Device-Centric Applications.}  
Another line of research focuses on embedding LLMs into everyday devices, including smartphones, portable translators, wearables, and extended reality (XR) systems. Wu et al.~\cite{wu2024smartphone} provide an overview of LLM-powered smartphones, covering market adoption, core functions, and risks. Wu et al.~\cite{wu2025assistantsadversariesexploringsecurity} extend this work with a security study of mobile LLM agents, identifying 11 distinct attack surfaces across reasoning, GUI interaction, and system-level execution. Beyond phones, Bohra et al.~\cite{bohra2025llm} propose \textit{SURYA-TAC}, an offline tactical translator for multilingual speech-to-speech communication in disconnected environments. Santos et al.~\cite{Santos2025Near} test LLM deployment on resource-constrained smartphones, showing feasibility but highlighting severe latency and memory costs.  
Wearables and XR devices represent another emerging frontier. Yousri et al.~\cite{yousri2024illusionxllmpoweredmixedreality} propose \textit{IllusionX}, a mixed-reality personal companion. Tang et al.~\cite{tang2025Integration} survey LLM integration in XR, while Wang et al.~\cite{wang2025surveylargelanguagemodels} categorize XR systems into technical paradigms and application domains. Concrete prototypes include haptic-feedback LLM glasses for navigation~\cite{tokmurziyev2025llmglassesgenaidrivenglasseshaptic}, hybrid smart glasses architectures that combine local and cloud services~\cite{grumeza2025hybrid}, and optimized smart home assistants at the edge~\cite{velaga2024optimizing}.

\textbf{Security and Privacy.}  
Security concerns form a critical research strand. Wu et al.~\cite{wu2025assistantsadversariesexploringsecurity} show that mobile agents are vulnerable to GUI manipulation and execution hijacking. Similar risks apply to XR and wearable systems, where multimodal sensing and continuous interaction enlarge the attack surface. Researchers call for secure-by-design frameworks, privacy-preserving deployment, and standardized defenses tailored to the unique properties of LLM-powered devices.

\subsubsection{Challenges \& Open Problems}

LLM-powered devices open up new ways of interacting with intelligent systems, but they also face a number of challenges that limit their practical adoption. These challenges can be summarized into three main aspects.

\underline{\textbf{Efficiency and Resource Constraints.}}  
Running LLMs on small, resource-limited devices is still very difficult. Smartphones, wearables, and XR headsets have strict limits on memory, storage, and battery life. Even when models are compressed or quantized, they often cannot provide responses quickly enough for smooth user interaction. Long response times, high memory consumption, and fast battery drain remain common problems. Moreover, many optimization methods that work well in cloud environments do not transfer easily to edge devices, because input patterns, workloads, and available hardware are very different. Achieving both high accuracy and real-time performance under such constraints is an unresolved problem.

\underline{\textbf{Security and Privacy Risks.}}  
Devices that integrate LLMs often process sensitive personal data such as voice, images, location, and system commands. This makes them vulnerable to both external attacks and misuse. On-device agents may be tricked by adversarial prompts, manipulated into unsafe actions, or exploited through system-level permissions. At the same time, continuous sensing of the user’s environment raises strong privacy concerns, as private information could be unintentionally recorded or leaked. Current solutions, such as local wake word detection or partial on-device processing, only address a fraction of the risks. Building secure and privacy-preserving device-level LLM systems remains an open challenge.

\underline{\textbf{System Integration and User Experience.}}  
LLM-powered devices rarely operate in isolation. They must coordinate speech recognition, vision understanding, haptic feedback, and sometimes cloud services, all while maintaining a seamless interaction loop. In practice, these pipelines are fragile. Small mismatches in latency across modules can break the flow of interaction, and many systems struggle to remain stable in noisy, dynamic, or unpredictable environments. Users often experience delays, inconsistent behavior, or confusing outputs, which weakens trust and limits adoption. To overcome this, future designs need not only better optimization but also new interaction paradigms that allow LLMs to adapt naturally to real-world conditions.

\begin{table}[htbp]
    \centering
    \caption{Challenges and Open Problems across LLM Application Paradigms.}
    \label{tab:llm_challenges}
    \resizebox{\linewidth}{!}{
    \begin{threeparttable}
    \begin{tabular}{lll}
        \toprule[1.2pt]
        \textbf{LLMApplication} & \textbf{Challenge} & \textbf{Description} \\
        \midrule[1.2pt]
        \multirow{3}{*}{LLM App Stores} 
        & Limited Flexibility & Fixed templates limit workflows, APIs, and advanced reasoning. \\
        & Platform Lock-in & Apps bound to store-specific rules, blocking reuse elsewhere. \\
        & Uncertain Quality Assurance & Weak or inconsistent reviews let risky apps pass through. \\
        \midrule
        \multirow{3}{*}{LLM Agents} 
        & Fragmented Protocols and Ecosystems & Frameworks use isolated APIs, hindering interoperability. \\
        & Lack of Process Visibility & Hard to link reasoning steps with actions and outcomes. \\
        & Security and Privacy Risks & Agents face prompt attacks, leaks, and covert collusion. \\
        \midrule
        \multirow{4}{*}{Self-hosted LLM Services} 
        & Efficiency and Resource Constraints & Latency, GPU memory, and energy remain bottlenecks. \\
        & Deployment Complexity & Full pipelines need heavy setup and maintenance. \\
        & Interoperability and Fragmentation & Incompatible APIs block migration and reuse. \\
        & Privacy and Governance & Risks from leaks, defaults, and weak compliance. \\
        \midrule
        \multirow{3}{*}{LLM-powered Devices} 
        & Efficiency and Resource Constraints & Devices lack memory, battery, and speed for real-time use. \\
        & Security and Privacy Risks & Sensitive multimodal data is easily exposed. \\
        & System Integration and User Experience & Speech, vision, and cloud modules fail to align smoothly. \\
        \bottomrule[1.2pt]
    \end{tabular}
    \end{threeparttable}}
\end{table}

\section{The Next Frontier of LLM Applications}
\label{sec:forntier}

\begin{figure}[htbp]
    \centering
    \includegraphics[width=1\linewidth]{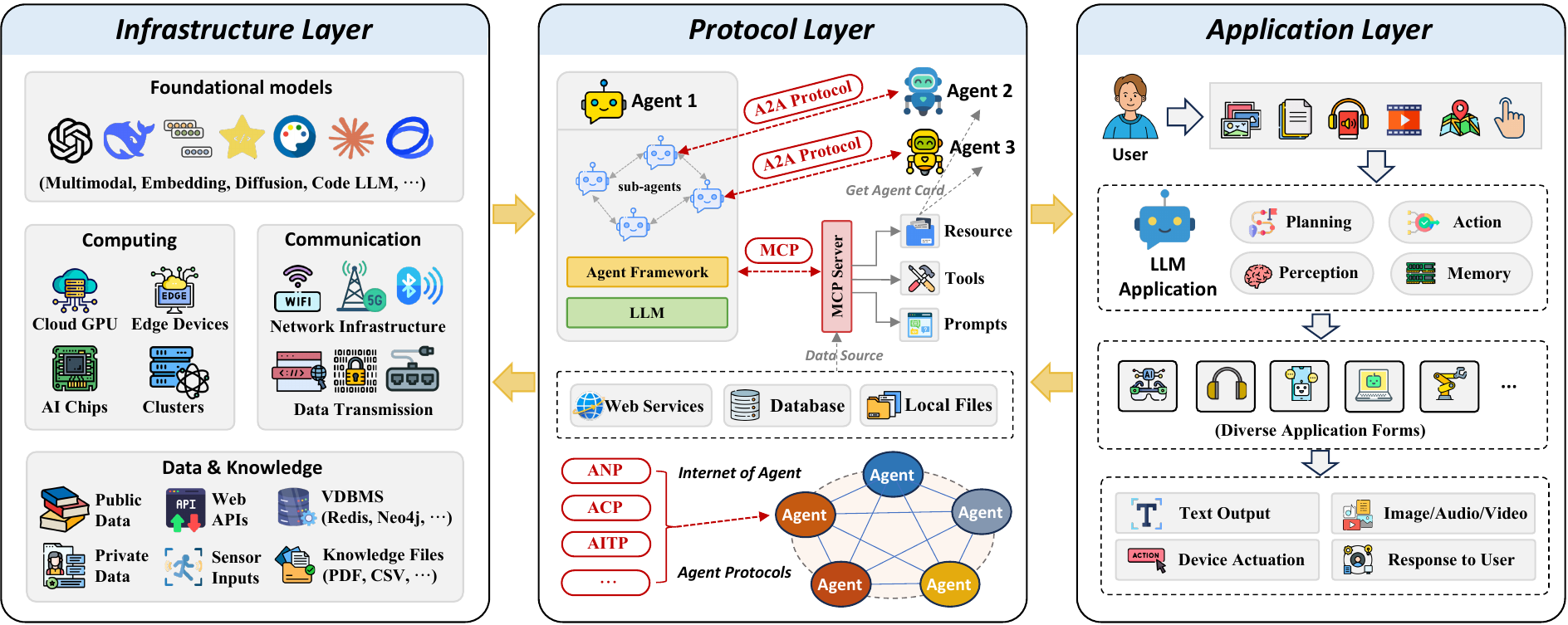}
    \caption{Three-layer architecture for the next frontier of LLM applications. The infrastructure layer provides foundational models, computing resources, communication networks, and data sources. The protocol layer coordinates interactions among agents, resources, and tools through standardized frameworks and protocols. The application layer delivers diverse application forms, where users interact with LLM-based systems for perception, planning, memory, and action.}
    \label{fig:frontier}
\end{figure}

The rapid evolution of LLM applications has revealed both their transformative potential and their fundamental limitations. As concluded in \autoref{tab:llm_challenges}, current LLM applications suffer from fragmented ecosystems, lack of process visibility, and persistent concerns about efficiency, security, and privacy. These issues constrain scalability, hinder interoperability, and erode user trust, especially as LLMs move beyond controlled cloud environments into self-hosted platforms, edge devices, and real-world deployments. 
To address these challenges, the next stage of development must go beyond incremental optimization. What is needed is a new design paradigm that integrates infrastructures, protocols, and applications into a unified architecture, as illustrated in \autoref{fig:frontier}. This vision of the \textit{Next Frontier of LLM Applications} emphasizes standardization, transparency, and trustworthy operation, enabling agents, services, and devices to collaborate seamlessly across diverse environments.

\subsection{Infrastructure Layer}  
The infrastructure layer gives the basic support for building and running LLM applications. The infrastructure layer is not a new invention but a systematic summary of the resources that current LLM applications already rely on. It brings together the essential components that serve as the energy carriers of intelligent systems. As LLM applications continue to take more diverse forms and provide more complex functions, these infrastructures remain the backbone that powers their operation. They determine how large a model can be deployed, how efficiently it can be served, and how securely it can interact with the environment. It has four main parts: foundational models, computing resources, communication systems, and data \& knowledge.  

\textbf{Foundational models} are the core. They supply the reasoning and generation ability. Multimodal LLMs handle text, image, and audio so that one model can understand mixed inputs. Embedding models turn queries or documents into vectors that support semantic search and RAG. Diffusion models create new images, audio, or video, which expands LLM use to creative tasks. Code LLMs write, debug, and explain programs, making them useful for software agents. These models are trained on large datasets and are reused across many applications.  
\textbf{Computing resources} make these models usable. Cloud GPUs and TPU clusters train and fine-tune LLMs with billions of parameters. They also support heavy inference workloads when latency is not strict. AI chips like NPUs, FPGAs, and edge accelerators allow models to run on local devices. Edge devices such as smartphones, smart glasses, and IoT boards give users access without depending fully on the cloud. The combination of cloud and edge resources enables both high performance and low latency in different settings.  
\textbf{Communication systems} connect models, agents, and users. WiFi 6 and 5G networks carry prompts and responses between devices and servers. Data transmission pipelines synchronize embeddings, documents, and results in real time. Network infrastructure also supports cross-device collaboration, where one agent running on an edge device can call tools or resources in the cloud. Reliable and low-latency communication is critical for interactive tasks such as voice assistants or robotics.  
\textbf{Data and knowledge} provide the grounding that LLMs need. Public data includes open corpora and web-scale text. Private data comes from company databases, user records, or local documents. Web APIs give access to live information such as weather or stock prices. Sensor inputs stream signals from cameras, microphones, or IoT devices. Knowledge files such as PDFs, CSVs, or manuals give structured references. Vector databases like Redis, Milvus, or Neo4j store embeddings so that queries can retrieve the most relevant context. This makes applications more reliable and domain-aware.

\subsection{Protocol Layer}
\label{subsec:protocol}

The protocol layer plays a central role in shaping the next generation of LLM applications. 
While the \textit{infrastructure layer} already provides the computational backbone and the \textit{application layer} defines diverse forms of intelligent services, 
it is the protocol layer that enables these components to interoperate in a scalable, secure, and standardized way. 
Without shared communication and coordination standards, LLM agents remain siloed, with each framework using bespoke interfaces and ad-hoc tool integrations. 
This fragmentation leads to duplicated engineering effort, limited interoperability, and security vulnerabilities. 
Standardized agent protocols address these challenges by offering structured methods for \textbf{tool invocation, context exchange, inter-agent communication, and governance}, 
paving the way for ecosystems of composable and trustworthy agents~\cite{yang2025surveyaiagentprotocols,ehtesham2025surveyagentinteroperabilityprotocols}. 

Recent years have seen the emergence of several key protocols that illustrate the potential of this layer. 
The \textbf{Model Context Protocol (MCP)}~\cite{anthropic2024mcp} defines a JSON-RPC based client-server model for secure access to external tools, resources, and prompts, 
thus providing a consistent interface for retrieval-augmented workflows and reducing integration overhead. 
The \textbf{Agent Communication Protocol (ACP)}~\cite{acp2025spec}, developed by IBM, supports MIME-typed multipart messages over REST and enables asynchronous, streaming-based communication at scale, 
making it suitable for heterogeneous enterprise environments. 
The \textbf{Agent-to-Agent Protocol (A2A)}~\cite{google2025a2a} provides peer-to-peer task delegation through \textit{Agent Cards} and structured message exchange, 
allowing specialized agents to collaborate securely across organizational boundaries. 
The \textbf{Agent Network Protocol (ANP)}~\cite{chang2024anp} extends this idea to the open web, enabling decentralized agent discovery and communication with W3C decentralized identifiers (DIDs) and JSON-LD based agent descriptions~\cite{ray2025review,derouiche2025agenticaiframeworksarchitectures,duan2025agentcommunicationsagenticai}. 
Together, these protocols begin to form the communication substrate of an emerging \textbf{Agentic Web}, where autonomous agents can discover, negotiate, and cooperate at scale~\cite{yang2025agenticwebweavingweb}. 

\begin{table}[h]
    \centering
    \caption{Representative Agent Protocols.}
    \label{tab:agent_protocols}
    \resizebox{0.98\linewidth}{!}{
    \begin{threeparttable}
    \begin{tabular}{lllll}
        \toprule[1.2pt]
        \textbf{Entity} & \textbf{Protocol} & \textbf{Proposer} & \textbf{Key Techniques} & \textbf{Core Problem Addressed} \\
        \midrule[1.2pt]
        \multirow{2}{*}{Context-Oriented} & MCP \cite{anthropic2024mcp} & Anthropic & RPC, OAuth & Standardized tool invocation and context sharing \\ \cmidrule{2-5}
                        & agent.json \cite{wildcardai2025agentsjson} & WildCardAI & .well-known & Unified agent discovery and metadata specification \\
        \midrule
        \multirow{7}{*}{Inter-Agent}   & A2A \cite{google2025a2a} & Google & RPC, OAuth & Secure peer-to-peer task delegation \\ \cmidrule{2-5}
                        & ANP \cite{chang2024anp} & Gaowei Chang & JSON-LD, DID & Decentralized discovery and collaboration \\ \cmidrule{2-5}
                        & AITP \cite{near2025aitp} & NEAR Foundation & Blockchain, HTTP & Trustworthy interaction and transaction logging \\ \cmidrule{2-5}
                        & AComP \cite{ibm2025acomp} & IBM / Linux Foundation AI & OpenAPI & General-purpose communication across agents \\ \cmidrule{2-5}
                        & AConP \cite{cisco2025aconp} & Cisco, LangChain & OpenAPI, JSON & Standardized service integration for agent workflows \\ \cmidrule{2-5}
                        & ACP \cite{acp2025spec} & IBM / Linux Foundation AI & RESTful HTTP, MIME & Structured session-aware messaging and routing \\ \cmidrule{2-5}
                        & Agora \cite{marro2024scalable} & Oxford University & Protocol Document & Scalable communication in large LLM networks \\ 
        \bottomrule[1.2pt]
    \end{tabular}
    \end{threeparttable}}
\end{table}

Despite these advances, the protocol layer is still at an early stage of development. 
Current implementations remain fragmented, and many lack mature governance, conformance testing, and wide adoption. 
Security guarantees, such as standardized identity verification, message signing, and auditability, are unevenly supported across protocols. 
Discovery and semantic interoperability remain open challenges: although mechanisms like \textit{Agent Cards} and \textit{Agent Description Protocols} exist, 
it is not yet clear how they will scale to support large, heterogeneous networks of agents with evolving capabilities. 
Moreover, existing protocols are often designed with either tool integration or agent-to-agent communication in mind, 
but not both, leaving open questions about how to integrate context-oriented and inter-agent protocols into a unified ecosystem. 

The protocol layer is a linchpin for the next frontier of LLM applications. 
By establishing common standards for communication, it can unlock interoperability, composability, and collective intelligence across diverse agents and devices. 
At the same time, its immaturity means there is vast space for further research, particularly in areas of \textbf{secure-by-design architectures, semantic alignment, decentralized discovery, and cross-protocol integration}.

\subsection{Application Layer}  
\label{subsec:application_layer}

The application layer is where users see and use LLM-powered systems. A user can give input through text, voice, images, or mixed formats. The application receives this input and processes it with modules for \textbf{planning, perception, action, and memory}. Planning decides the next steps. Perception reads and interprets input. Action calls tools, services, or devices. Memory keeps track of past context and user data.
This layer connects with the \textbf{infrastructure layer}. It uses compute, storage, and communication to run. It also connects with the \textbf{protocol layer}, which provides standards for tool use, data sharing, and multi-agent communication. Protocols such as \textit{MCP}, \textit{A2A}, and \textit{ANP} help different agents and applications talk to each other. Without them, applications stay isolated and cannot work across platforms.

LLM applications can take many forms. Some are text-based chatbots. Others are multimodal assistants that combine speech, video, and image generation. Some can control external systems, such as \textbf{IoT devices}, \textbf{smart homes}, or \textbf{industrial robots}. New hardware such as \textbf{AI smartphones}, \textbf{smart glasses}, and \textbf{wearable devices} bring LLM functions closer to daily life. In these devices, agents can capture sensor signals, gestures, and environmental data, and then give feedback through text, voice, haptics, or visual display.
More advanced systems show the trend of \textbf{embodied intelligence}. Here, an LLM agent is not only a software program but also part of a physical device. For example, a robot with an LLM can plan paths, respond to commands, and act in the real world. A car with an LLM can analyze voice, images, and driving data at the same time. A wearable assistant can listen, watch, and give advice instantly. These forms show how applications will expand from screen-based chat to real-time, interactive, and embodied services.

The application layer must also become \textbf{open and interoperable}. A single vendor app is no longer enough. An LLM assistant in a phone should work with home devices, office tools, or other agents. Protocol support such as \textit{MCP} or \textit{A2A} makes this possible. Applications need modular design so that developers can extend them with new functions, plug-ins, or device links. At the same time, \textbf{security and privacy} should be included from the start. Data exchange, memory use, and device control must be safe to avoid leaks or abuse.

\section{Future Directions}
\label{sec:future}

As LLM applications evolve toward more autonomous, adaptive, and interconnected systems, several critical challenges remain unresolved, while at the same time new opportunities are emerging. In this section, we outline the key \textbf{challenges} and highlight concrete \textbf{opportunities} for the next frontier of LLM applications. This perspective emphasizes the interplay between infrastructure, protocol, and application layers, with a focus on building secure, interoperable, and human-aligned ecosystems.

\subsection{Challenges}  

\subsubsection{Protocol Fragmentation and Limited Interoperability}  
LLM agents today are built inside isolated ecosystems. Each framework or platform defines its own method for tool invocation, memory management, and data exchange, which makes it difficult to connect agents developed under different environments. For example, some frameworks rely on JSON-based schemas while others use custom RPC or plugin formats, and context persistence mechanisms vary widely. This heterogeneity prevents seamless transfer of state and capabilities across systems.  
Even when cross-framework communication is technically possible, there is no common agreement on discovery mechanisms, authentication, or negotiation rules. Without these, agents cannot trust or identify each other automatically, which increases the cost of coordination in multi-agent workflows. This limits scalability for applications that need to combine capabilities from different platforms or organizations.  
Early attempts such as MCP, A2A, and ANP show potential steps toward interoperability by defining shared message formats and interaction protocols, yet adoption is fragmented. Cross-ecosystem integration remains rare, and without broader standardization, the ecosystem risks evolving into isolated silos rather than a connected agent network.

\subsubsection{Security Risks in Open Agent Environments}  

The move toward open and interoperable LLM applications creates not only opportunities for richer functionality but also a growing set of security risks. When agents can freely connect with external tools, invoke APIs, and collaborate across platforms, each interaction point becomes a potential attack surface. These risks are qualitatively different from those in closed application stores, because execution environments are no longer tightly controlled, and agents operate in dynamic, multi-party settings.  

One class of threats arises in the \textbf{tool calling process}. LLM applications that dynamically compose workflows or execute external commands are vulnerable to injection-style attacks. For example, indirect prompt injection can exploit data sources to manipulate downstream tool execution~\cite{greshake2023not,pedro2023from}, while recent studies show that adversarial queries can even extract hidden system prompts from deployed applications~\cite{hui2024pleak}. These attacks blur the boundary between natural language inputs and executable instructions, making it difficult to guarantee integrity once agents operate in open environments.  
A second risk lies in \textbf{persistent agent memory and context sharing}. Memory modules are increasingly used to store private user data or long-term interaction histories. However, memory extraction attacks demonstrate that adversaries can systematically recover sensitive information by carefully crafted queries~\cite{wang2025unveiling}. In multi-agent systems, such leaks are not confined to a single agent; compromised memory can be propagated to collaborators, amplifying the scale of the breach.  
Finally, \textbf{multi-agent collaboration introduces systemic vulnerabilities}. When agents coordinate via shared protocols, malicious participants can collude or spread misinformation across the network. Prior work highlights the possibility of covert communication between agents through steganographic or indirect channels~\cite{motwani2024advances,peignelefebvre2025multiagent}, which undermines monitoring and governance mechanisms. Moreover, weak privilege management can allow attackers to escalate control or maintain unauthorized access beyond the intended session lifecycle~\cite{liu2024demystifying,tete2024threat}.  

These examples illustrate how opening LLM applications to a broader ecosystem magnifies risks across the entire lifecycle of interaction, from tool invocation to long-term coordination. Traditional safeguards such as input filtering or static access control are not sufficient in such fluid environments, underscoring the urgent need for secure-by-design protocols and continuous monitoring of agent-tool interactions.

\subsubsection{Lack of Process Visibility and Accountability}  
Users and developers often see only the input and final output of LLM applications, while the intermediate reasoning traces, memory updates, and inter-agent communication remain hidden. This creates a ``black box'' workflow where the internal decision path cannot be inspected or verified. For example, when an output is incorrect, there is no clear trace showing whether the error comes from flawed reasoning, missing context, or unreliable external calls. Similarly, when multiple agents interact, it is often impossible to reconstruct what information was shared or how coordination decisions were reached.  
This lack of visibility complicates tasks such as debugging, auditing, and performance monitoring. Developers cannot easily identify bottlenecks or misconfigurations, and users cannot verify that the system has respected constraints such as data privacy or safety rules. Moreover, accountability becomes a critical concern: when an autonomous agent executes actions on behalf of a user, there is no transparent link between the reasoning steps and the external behavior. In high-stakes domains such as healthcare, finance, or government services, this opacity raises serious questions about trust, compliance, and liability.

\subsubsection{Resource and Deployment Constraints}  
Running LLM-powered applications still requires large amounts of compute, memory, and energy. On the cloud, costs are high and latency is hard to control. On edge or device-level deployments, models are constrained by hardware limits, which reduces their accuracy and robustness. Techniques such as quantization, pruning, and distillation help reduce resource demands, but they often degrade performance. Moreover, deployment pipelines require integrating serving engines, embeddings, vector databases, and monitoring systems, which makes them complex and error-prone.  

\subsubsection{Closed Ecosystems and Limited Extensibility}  
Many emerging LLM applications are still designed as closed ecosystems, where workflows are predefined and connectors to external systems are proprietary. Such designs limit extensibility: developers cannot easily replace components, incorporate third‑party modules, or adapt workflows to new contexts. This restricts innovation at the application layer and makes it difficult to reuse capabilities across projects.  For next‑frontier applications, openness is essential. Future ecosystems will require shared protocols for agent interaction, modular workflow definitions, and plug‑and‑play integration with heterogeneous devices. Without these, applications will remain siloed, unable to exploit the full potential of embodied intelligence scenarios where LLMs collaborate with robots in manufacturing, interface with AR glasses in education, or connect with smart home systems in consumer environments. The absence of extensibility thus poses a fundamental barrier to scaling from isolated agents toward multi‑modal, real‑world intelligent assistants.

\subsection{Opportunities}  

\subsubsection{Protocol-Driven Ecosystems}  
A realistic opportunity for achieving a more unified and open LLM application ecosystem lies in the development of shared protocols. These protocols do not need to reinvent the foundation of distributed systems; rather, they can extend well-established web technologies such as HTTP, gRPC, JSON schemas, and OAuth-based authentication to the LLM domain. For example, standardized model invocation protocols could define common request and response formats for prompt submission, tool calling, and result streaming, ensuring that applications can interoperate across frameworks.  

Beyond communication, openness requires protocols for representing workflows, reasoning states, and long-term memory. Portable formats for context and vector indexes would make it possible to migrate applications across frameworks without losing history or auditability. Similarly, shared conventions for reporting latency, token usage, and system load would help orchestrators distribute requests more efficiently across heterogeneous infrastructures. Looking ahead, standard bindings to robotics middleware such as ROS2, IoT protocols like MQTT, or AR/VR streaming interfaces would also be necessary to support embodied intelligence scenarios.  

These protocol layers would serve as the connective tissue of the ecosystem. By relying on standards for invocation, discovery, authentication, workflow representation, and device integration, future LLM applications could move beyond isolated silos and evolve toward an open, interoperable, and extensible environment where agents, models, and tools can be freely combined.

\subsubsection{Secure by Design}   

Security in LLM applications should not come after deployment. It should be included in the design process from the start. It shapes how applications, protocols, and infrastructures are built and deployed. A secure-by-design approach begins with defining what needs protection, such as user data, model parameters, and interaction logs. It also sets a clear threat model that considers attackers like malicious users, compromised tools, and colluding agents in multi-agent systems. By doing this early, system architects can align functionality with core security principles like least privilege, provenance, and auditability.  

\textbf{At the application layer}, security by design means that each component (planning, memory, perception, and action) is built with protection in mind. Tool calls pass through gateways that use allowlists and schema checks. Memory modules separate temporary storage from permanent storage, use encryption, and apply strict expiration rules. Retrieval-augmented generation pipelines check sources and clean inputs to avoid poisoning. Prompts are built from structured templates, not free text, to reduce risks of injection. Reasoning traces and actions are tied together through shared logs so developers can track decisions and verify agent behavior.  
\textbf{At the protocol layer}, secure communication depends on identity management, authenticated sessions, and clear capability rules. Protocols like MCP or A2A can be extended with session tokens, typed permissions, and consent markers to control agent and tool interaction. Standard schemas and message checks protect against malformed or harmful exchanges. Reputation systems and quarantine methods help contain agents that behave badly.  
\textbf{At the infrastructure layer}, confidential computing, secure enclaves, and encrypted vector stores keep models and embeddings safe. Signed models and datasets, along with reproducible build pipelines, help prevent backdoors. Containers and sandboxes isolate tools or plugins to reduce harm if one is compromised.  

Secure-by-design also applies to the development cycle. Continuous integration should include security tests with adversarial prompts, memory extraction checks, and tool misuse scenarios. Red-teaming, staged releases, and anomaly detection can give added protection. For users, defaults should protect privacy with minimal data collection, local pre-processing, and short data retention. Clear options should let users inspect, delete, or export their data.  
In multi-agent systems, agents should be treated as untrusted by default. Each agent should have isolated state and limited capabilities. Important decisions should require approvals from more than one agent or cross-checks across diverse models. Monitoring tools should detect hidden communication channels and enforce standard content formats.

\subsubsection{Human-Centered Monitoring and Control}  
The lack of transparency in agent behavior can be turned into an opportunity for richer interaction design. Developers can build monitoring dashboards that visualize each reasoning step, highlight decision points, and provide options for user intervention. Enterprises can deploy compliance panels where managers set policies that agents must follow (e.g., blocking unsafe actions). Consumers can use mobile apps to view, edit, or rewind agent actions, similar to browsing browser history. These mechanisms not only improve trust but also differentiate products by offering explainability as a feature.  

\subsubsection{Device Integration and Ubiquity}  
The miniaturization of AI hardware creates a concrete opportunity to embed LLMs into everyday devices and environments. For instance, smart earbuds can run local speech-to-text models that transcribe conversations in real time and perform short-form summarization without relying on the cloud. AR glasses can host lightweight reasoning modules that support instant translation of street signs or context-aware navigation assistance during walking or driving. Smart home robots can combine on-device processing with cloud agents to execute voice commands, optimize household schedules, or coordinate with other appliances.  
These deployments become more practical as hardware vendors integrate neural processing units (NPUs), tensor accelerators, and energy‑efficient GPUs into consumer devices such as smartphones or wearable headsets. At the same time, protocol standards for communication and discovery allow these devices to interoperate instead of functioning as isolated silos. The result is a scalable ecosystem of embodied agents, where LLM-powered capabilities are not confined to single applications but pervasively embedded into daily activities through a network of smart and connected devices.

\subsubsection{Composable and Extensible Applications}  
The next wave of LLM applications can move away from monolithic agents and toward modular architectures. In this model, developers release micro-capabilities, such as code execution, scheduling, search, or retrieval, that can be composed dynamically into workflows. A protocol-based interoperability layer would ensure that these micro-capabilities are portable and can be reused across platforms, rather than being locked into a single framework. For enterprises, this could resemble a ``plugin marketplace'' where teams assemble domain-specific agents by selecting validated modules instead of building entire systems from scratch.  
Several research directions can make this vision feasible. One line is the design of \textbf{standard invocation schemas and workflow languages} that describe how modules interact, similar to open APIs for services but specialized for agent reasoning. Another is the development of \textbf{safe composition mechanisms}, for example by isolating execution in sandboxes, handling conflicts between modules automatically, and enforcing resource limits. A third direction is the creation of \textbf{cross-platform packaging and distribution standards}, which would enable modules to be discovered, installed, and updated consistently across ecosystems. Together, these directions would lower development costs, speed up innovation, and allow user-level customization, ultimately making LLM ecosystems more flexible and sustainable.

\section{Limitation}
This paper focuses on highlighting the current landscape and future directions of LLM applications. The four paradigms we adopt (LLM app stores, LLM agents, self-hosted services, and LLM-powered devices) are not meant to be exhaustive. We chose these four paradigms because they appear most often in recent papers and industrial platforms on LLM applications. They reflect how users today access and deploy LLMs in practice. This division is meant to capture common patterns, not to claim a definitive taxonomy.
Second, our discussion of related research emphasizes emerging trends instead of offering a systematic or comprehensive review of the literature. We do not claim to cover all existing work; instead, we highlight representative directions that help illustrate challenges and opportunities for the next frontier. 
These choices inevitably leave out other relevant paradigms and studies, which call for further exploration in future research.

\section{Conclusion}
\label{sec:conclusion}

This paper has provided a structured review of current LLM application paradigms: LLM app stores, LLM agents, self-hosted services, and LLM-powered devices, highlighting their architectures, ecosystems, and open problems. We argued that the next frontier cannot be achieved by incremental improvements alone, but requires a systematic architecture consisting of infrastructure, protocol, and application layers. Within this framework, we identified the major challenges of fragmentation, security, visibility, resource constraints, and limited extensibility, and outlined concrete opportunities such as protocol-driven ecosystems, secure-by-design practices, human-centered monitoring, device ubiquity, and composable applications. The overall direction points toward open, trustworthy, and sustainable LLM ecosystems that evolve from siloed solutions into reliable foundations for everyday intelligent services.

\bibliographystyle{ACM-Reference-Format}
\bibliography{acmart}

\end{document}